\documentclass[aps,twocolumn,superscriptaddress]{revtex4}
\usepackage{epsfig}
\usepackage{times}
\begin{document}

\title{Directional learning and the provisioning of public goods}

\author{Heinrich H. Nax}
\email{hnax@ethz.ch}
\affiliation{Department of Social Sciences, ETH Z\"{u}rich, Clausiusstrasse 37-C3, 8092 Zurich, Switzerland}

\author{Matja{\v z} Perc}
\email{matjaz.perc@uni-mb.si}
\affiliation{Faculty of Natural Sciences and Mathematics, University of Maribor, Koro{\v s}ka cesta 160, SI-2000 Maribor, Slovenia}
\affiliation{Faculty of Science, King Abdulaziz University, Jeddah, Saudi Arabia}

\begin{abstract}
We consider an environment where players are involved in a public goods game and must decide repeatedly whether to make an individual contribution or not. However, players lack strategically relevant information about the game and about the other players in the population. The resulting behavior of players is completely uncoupled from such information, and the individual strategy adjustment dynamics are driven only by reinforcement feedbacks from each player's own past. We show that the resulting ``directional learning'' is sufficient to explain cooperative deviations away from the Nash equilibrium. We introduce the concept of $k-$strong equilibria, which nest both the Nash equilibrium and the Aumann-strong equilibrium as two special cases, and we show that, together with the parameters of the learning model, the maximal $k-$strength of equilibrium determines the stationary distribution. The provisioning of public goods can be secured even under adverse conditions, as long as players are sufficiently responsive to the changes in their own payoffs and adjust their actions accordingly. Substantial levels of public cooperation can thus be explained without arguments involving selflessness or social preferences, solely on the basis of uncoordinated directional (mis)learning.
\end{abstract}

\maketitle

Cooperation in sizable groups has been identified as one of the pillars of our remarkable evolutionary success. While between-group conflicts and the necessity for alloparental care are often cited as the likely sources of the other-regarding abilities of the genus \textit{Homo} \cite{bowles_11, hrdy_11}, it is still debated what made us the ``supercooperators'' that we are today \cite{nowak_11, rand_tcs13}. Research in the realm of evolutionary game theory \cite{maynard_82, weibull_95, hofbauer_98, mestertong_01, nowak_06, myatt_res08} has identified a number of different mechanisms by means of which cooperation might be promoted \cite{mestertong_qrb92, nowak_s06}, ranging from different types of reciprocity and group selection to positive interactions \cite{rand_s09}, risk of collective failure \cite{santos_pnas11}, and static network structure \cite{santos_n08, rand_pnas14}.

The public goods game \cite{Isaac_85}, in particular, is established as an archetypical context that succinctly captures the social dilemma that may result from a conflict between group interest and individual interests \cite{ledyard_97, chaudhuri_ee11}. In its simplest form, the game requires that players decide whether to contribute to a common pool or not. Regardless of the chosen strategy by the player himself, he receives an equal share of the public good which results from total contributions being multiplied by a fixed rate of return. For typical rates of return it is the case that, while the individual temptation is to free-ride on the contributions of the other players, it is in the interest of the collective for everyone to contribute. Without additional mechanisms such as punishment \cite{fehr_aer00}, contribution decisions in such situations \cite{ledyard_97, chaudhuri_ee11} approach the free-riding Nash equilibrium \cite{nash_pnas50} over time and thus lead to a ``tragedy of the commons'' \cite{hardin_g_s68}. Nevertheless, there is rich experimental evidence that the contributions are sensitive to the rate of return \cite{fischbacher_el01} and positive interactions \cite{rand_s09}, and there is evidence in favor of the fact that social preferences and beliefs about other players' decisions are at the heart of individual decisions in public goods environments \cite{fischbacher_aer10}.

In this paper, however, we shall consider an environment where players have no strategically relevant information about the game and/ or about other players, and hence explanations in terms of social preferences and beliefs are not germane. Instead, we shall propose a simple learning model, where players may mutually reinforce learning off the equilibrium path. As we will show, this phenomenon provides an alternative and simple explanation for why contributions rise with the rate of return, as well as why, even under adverse conditions, public cooperation may still prevail. Previous explanations of this experimental regularity \cite{ledyard_97} are based on individual-level costs of `error' \cite{palfrey_97, goeree_05}.

Suppose each player knows neither who the other players are, nor what they earn, nor how many there are, nor what they do, nor what they did, nor what the rate of return of the underlying public goods game is. Players do not even know whether the underlying rate of return stays constant over time (even though in reality it does) because their own payoffs are changing due to the strategy adjustments of other players, about which they have no information. Without any such knowledge, players are unable to determine ex ante whether contributing or not contributing is the better strategy in any given period, i.e., players have no strategically relevant information about how to respond best. As a result, the behavior of players has to be \emph{completely uncoupled} \cite{foster_te06, young_geb09}, and their strategy adjustment dynamics are likely to follow a form of \emph{reinforcement} \cite{roth_geb95, erev_aer98} feedback or, as we shall call it, \emph{directional learning} \cite{selten_86, selten_94}. We note that, in our model, due to the one-dimensionality of the strategy space, reinforcement and directional learning are both adequate terminologies for our learning model. Since reinforcement applies also to general strategy spaces and is therefore more general we will prefer the terminology of directional learning. Indeed, such directional learning behavior has been observed in recent public goods experiments \cite{bayer_13, young_13}. The important question is how \textit{well} will the population learn to play the public goods game despite the lack of strategically relevant information. Note that \textit{well} here has two meanings due to the conflict between private and collective interests: on the one hand, how close will the population get to playing the Nash equilibrium, and, on the other hand, how close will the population get to playing the socially desirable outcome.

The learning model considered in this paper is based on a particularly simple ``directional learning'' algorithm which we shall now explain. Suppose each player plays both cooperation (contributing to the common pool) and defection (not contributing) with a mixed strategy and updates the weights for the two strategies based on their relative performances in previous rounds of the game. In particular, a player will increase its weight on contributing if a previous-round switch from not contributing to contributing led to a higher realized payoff or if a previous-round switch from contributing to not contributing led to a lower realized payoff. Similarly, a player will decrease its weight on contributing if a previous-round switch from contributing to not contributing led to a higher realized payoff or if a previous-round switch from not contributing to contributing led to a lower realized payoff. For simplicity, we assume that players make these adjustments at a fixed incremental step size $\delta$, even though this could easily be generalized. In essence, each player adjusts its mixed strategy directionally depending on a Markovian performance assessment of whether a previous-round contribution increase/decrease led to a higher/lower payoff.

Since the mixed strategy weights represent a well-ordered strategy set, the resulting model is related to the directional learning/ aspiration adjustment models \cite{sauermann_62, selten_86, selten_94}, and similar models have previously been proposed for bid adjustments in assignment games \cite{nax_ieee13}, as well as in two-player games \cite{laslier_14}. In \cite{nax_ieee13} the dynamic leads to stable cooperative outcomes that maximize total payoffs, while Nash equilibria are reached in \cite{laslier_14}. The crucial difference between these previous studies and our present study is that our model involves more than two players in a voluntary contributions setting, and, as a result, that there can be interdependent directional adjustments of groups of players including more than one but not all the players. This can lead to uncoordinated (mis)learning of subpopulations in the game.

Consider the following example. Suppose all players in a large standard public goods game do not contribute to start with. Then suppose that a player in a subpopulation uncoordinatedly but by chance simultaneously decide to contribute. If this group is sufficiently large (the size of which depends on the rate of return), then this will result in higher payoffs for all players including the contributors, despite the fact that not contributing is the dominant strategy in terms of unilateral replies. In our model, if indeed this generates higher payoffs for all players including the freshly-turned contributors, then the freshly-turned contributors would continue to increase their probability to contribute and thus increase the probability to trigger a form of stampede or herding effect, which may thus lead away from the Nash equilibrium and towards a socially more beneficial outcome.

Our model of uncoordinated but mutually reinforcing deviations away from Nash provides an alternative explanation for the following regularity that has been noted in experiments on public goods provision \cite{ledyard_97}. Namely, aggregate contribution levels are higher the higher the rate of return, despite the fact that the Nash equilibrium remains unchanged (at no-contribution). This regularity has previously been explained only at an individual level, namely that `errors' are less costly -- and therefore more likely -- the higher the rate of return, following quantal-response equilibrium arguments \cite{palfrey_97, goeree_05}. By contrast, we provide a group-dynamic argument. Note that the alternative explanation in terms of individual costs is not germane in our setting, because we have assumed that players have no information to make such assessments. It is in this sense that our explanation perfectly complements the explanation in terms of costs.

In what follows, we present the results, where we first set up the model and then deliver our main conclusions. We discuss the implications of our results in section 3. Further details about the applied methodology are provided in the Methods section.

\section*{Results}

\subsection*{Public goods game with directional learning}
In the public goods game, each player $i$ in the population $N={1,2,...,n}$ chooses whether to contribute ($c_i=1$) or not to contribute ($c_i=0$) to the common pool. Given a fixed \emph{rate of return} $r>0$, the resulting payoff of player $i$ is then $u_i=(1-c_i )+(r /n)* \sum_{j\in N} c_j$. We shall call $r/n$ the game's \emph{marginal per-capita rate of return} and denote it as $R$. Note that for simplicity, but without loss of generality, we have assumed that the group is the whole population. In the absence of restrictions on the interaction range of players \cite{perc_jrsi13}, i.e., in well-mixed populations, the size of the groups and their formation can be shown to be of no relevance in our case, as long as $R$ rather than $r$ is considered as the effective rate of return.

The directional learning dynamics is implemented as follows. Suppose the above game is infinitely repeated at time steps $t=0,1,2,...$, and suppose further that $i$, at time $t$, plays $c^t_i=1$ with probability $p_i^t\in [\delta,1-\delta]$ and $c^t_i=0$ with probability $(1-p_i^t)$. Let the vector of contribution probabilities $p^t$ describe the state of the game at time $t$. We initiate the game with all $p_i^0$ lying on the $\delta$-grid between $0$ and $1$, while subsequently individual mixed strategies evolve randomly subject to the following three ``directional bias'' rules:
\begin{description}
\item[upward:] if $u_i (c_i^t )>u_i (c_i^{t-1})$ and $c_i^t>c_i^{t-1}$,
or if $u_i (c_i^t )<u_i (c_i^{t-1})$ and $c_i^t<c_i^{t-1}$,
then $p_i^{t+1}=p_i^t+\delta$ if $p_i^t<1$; otherwise, $p_i^{t+1}=p_i^t$.
\item[neutral:] if $u_i (c_i^t )=u_i (c_i^{t-1})$ and/or $c_i^t=c_i^{t-1}$,
then $p_i^{t+1}=p_i^t$ , $p_i^{t}+\delta$, or $p_i^t-\delta$ with equal probability if $0<p_i^t<1$; otherwise, $p_i^{t+1}=p_i^t$.
\item[downward:] if $u_i (c_i^t )>u_i (c_i^{t-1})$ and $c_i^t<c_i^{t-1}$,
or if $u_i (c_i^t )<u_i (c_i^{t-1})$ and $c_i^t>c_i^{t-1}$,
then $p_i^{t+1}=p_i^t-\delta$ if $p_i^t>0$; otherwise, $p_i^{t+1}=p_i^t$.
\end{description}

Note that the second, neutral rule above allows random deviations from any intermediate probability $0<p_i<1$. However, $p_i=0$ and $p_i=1$ for all $i$ are absorbing state candidates. We therefore introduce perturbations to this directional learning dynamics and study the resulting stationary states. In particular, we consider perturbations of order $\epsilon$ such that, with probability $1-\epsilon$, the dynamics is governed by the original three ``directional bias'' rules. However, with probability $\epsilon$, either $p_i^{t+1}=p_i^t$, $p_i^{t+1}=p_i^t-\delta$ or $p_i^{t+1}=p_i^t+\delta$ happens equally likely (with probability $\epsilon/3$) but of course obeying the $p_i^{t+1} \in[0,1]$ restriction.

\subsection*{Provisioning of public goods}
We begin with a formal definition of the \emph{$k-$strong equilibrium}. In particular, a pure strategy imputation $s^*$ is a $k$-strong equilibrium of our (symmetric) public goods game if, for all $C\subseteq N$ with $|C|\leq k$, $u_i (s_C^*;s_{N\setminus C}^*) \geq  u_i (s_C'; s_{N\setminus C}^*) $ for all $i\in C$ for any alternative pure strategy set $s_C'$ for $C$. As noted in the previous section, this definition bridges, one the one hand, the concept of the Nash equilibrium in pure strategies \cite{nash_pnas50} in the sense that any $k-$strong equilibrium with $k>0$ is also a Nash equilibrium, and, on the other hand, that of the (Aumann-)strong equilibrium \cite{aumann_74, aumann_87} in the sense that any $k-$strong equilibrium with $k=n$ is Aumann strong. Equilibria in between (for $1<k<n$) are ``more stable'' than a Nash equilibrium, but ``less stable'' than an Aumann-strong equilibrium.

\begin{figure}
\centerline{\epsfig{file=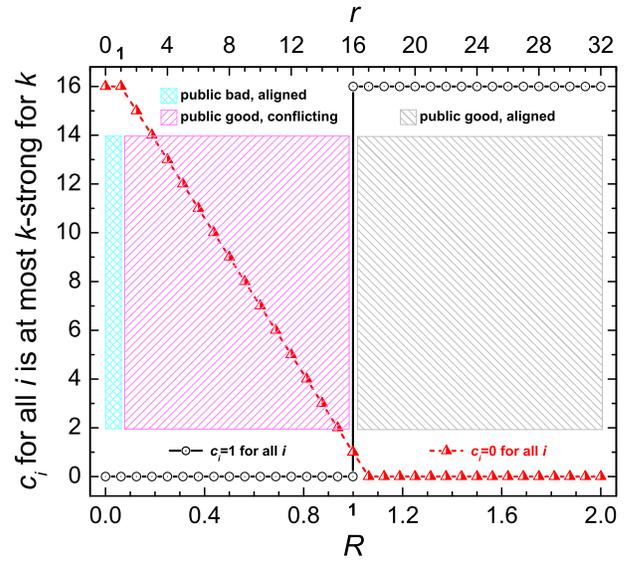,width=8.24cm}}
\caption{\label{ana} The maximal $k$-strength of equilibria in the studied public goods game with directional learning. As an example, we consider the population size being $n=16$. As the rate of return $r$ increases above $1$, the Aumann-strong ($n-$strong) $c_i=0$ for all $i$ (full defection) equilibrium looses strength. It is still the unique Nash equilibrium, but its maximal strength is bounded by $k=17-r$. As the rate of return $r$ increases further above $n$ ($R>1$), the $c_i=1$ for all $i$ (full cooperation) equilibrium suddenly becomes Aumann-strong ($n-$strong). Shaded regions denote the public bad game ($r<1$), and the public goods games with conflicting ($1<r<n$) and aligned ($R>1$) individual and public motives in terms of the Nash equilibrium of the game (see main text for details). We note that results for other population and/or group sizes are the same over $R$, while $r$ and the slope of the red line of course scale accordingly.}
\end{figure}

The maximal $k$-strengths of the equilibria in our public goods game as a function of $r$ are depicted in Fig.~\ref{ana} for $n=16$. The cyan-shaded region indicates the ``public bad game'' region for $r<1$ ($R<1/n$), where the individual and the public motives in terms of the Nash equilibrium of the game are aligned towards defection. Here $c_i=0$ for all $i$ is the unique Aumann-strong equilibrium, or in terms of the definition of the $k-$strong equilibrium, $c_i=0$ for all $i$ is $k-$strong for all $k\in[1,n]$. The magenta-shaded region indicates the typical public goods game for $1<r<n$ ($1/n<R<1$), where individual and public motives are conflicting. Here there exists no Aumann-strong equilibria. The outcome $c_i=0$ for all $i$ is the unique Nash equilibrium, and that outcome is also $k$-strong equilibrium for some $k\in [1,n)$, where the size of $k$ depends on $r$ and $n$ in that $\partial k/\partial r \leq 0$ while $\partial k/\partial n \geq 0 $. Finally, the gray-shaded region indicates the unconflicted public goods game for $r>n$ ($R>1$), where individual and public motives are again aligned, but this time towards cooperation. Here $c_i=1$ for all $i$ abruptly becomes the unique Nash and Aumann-strong equilibrium, or equivalently the unique $k-$strong equilibrium for all $k\in [1,n]$.

If we add perturbations of order $\epsilon$ to the unperturbed public goods game with directional learning that we have introduced in section 2, there exist stationary distributions of $p_i$ and the following proposition can be proven. In the following, we denote by ``$k$'' the maximal $k-$strength of an equilibrium.

\begin{description}
\item [Proposition:] As $t\rightarrow \infty$, starting at any $p^0$, the expectation with respect to the stationary distribution is $E[p^t]> 1/2$ if $ R\geq 1$ and $E[p^t]<1/2$ if $R<1$. $\partial E[p^t ]/\partial \epsilon<0$ if $R\geq 1$, and $\partial E[p^t ]/\partial \epsilon>0$ if $R< 1$. Moreover, $\partial E[p^t ]/\partial \delta>0$, and $\partial E[p^t ]/ \partial \delta<0$ if $R\geq 1$. Finally, $\partial E[p^t ]/ \partial k<0$ if $R<1$.
\end{description}

We begin the proof by noting that the perturbed process given by our dynamics results in an irreducible and aperiodic Markov chain, which has a unique stationary distribution. When $\epsilon=0$, any absorbing state must have $p_i^t=0 $ or $1$ for all players. This is clear from the positive probability paths to either extreme from intermediate states given by the unperturbed dynamics. We shall now analyze whether $p_i^t=0$ or $1$, given that $p_j^t=0$ or $1$ for all $j\neq i$, has a larger attraction given the model's underlying parameters.

If $R\geq 1$, the probability path for any player to move from $ p_i^t=0 $ to $p_i^{t+T}=1$ in some $T=1/\delta$ steps requires a single perturbation for that player and is therefore of the order of a single $\epsilon$. By contrast, the probability for any player to move from $p_i^t=1$ to $p_i^{t+T}=0$ in $T$ steps is of the order $\epsilon^3$, because at least two other players must increase their contribution in order for that player to experience a payoff increase from his non-contribution. Along any other path or if $p^t$ is such that there are not two players $j$ with $p_j^t=0$ to make this move, then the probability for $i$ to move from $p_i^t=1$ to $p_i^{t+T}=0$ in $T$ steps requires even more perturbations and is of higher order. Notice that, for any one player to move from $p_i^t=0 $ to $p_i^{t+T}=1$ we need at least two players to move away from $p_i^t=0$ along the least-resistance paths. Because contributing 1 is a best reply for all $R\geq 1$, those two players will also continue to increase if continuing to contribute 1. Notice that the length of the path is $T=1/ \delta$ steps, and that the path requires no perturbations along the way, which is less likely the smaller $\delta$.

\begin{figure}
\centerline{\epsfig{file=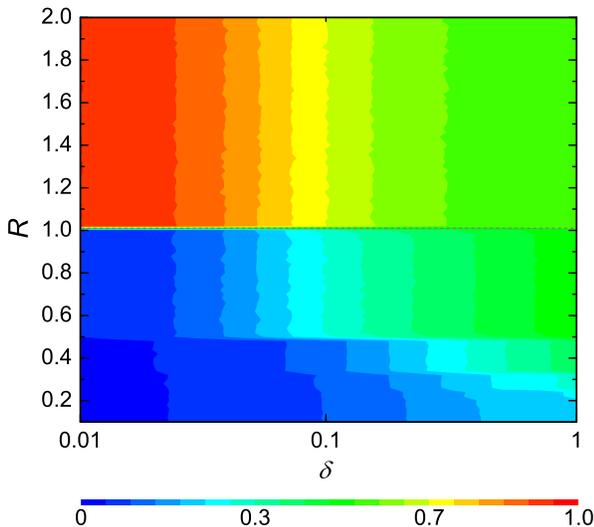,width=8cm}}
\caption{\label{simul} Color-encoded average contribution levels in the unperturbed public goods game with directional learning. Simulations confirm that, with little directional learning sensitivity (i.e. when $\delta$ is zero or very small), for the marginal per-capita rate of return $R>1$ the outcome $c_i=1$ for all $i$ is the unique Nash and Aumann-strong equilibrium. For $R=1$ (dashed horizontal line), any outcome is a Nash equilibrium, but only $c_i=1$ for all $i$ is Aumann-strong while all other outcomes are only Nash equilibria. For $R<1$, $c_i=0$ for all $i$ is the unique Nash equilibrium, and its maximal $k-$strength depends on the population size. This is in agreement with results presented in Fig.~\ref{ana}. Importantly, however, as the responsiveness of players increases, contributions to the common pool become significant even in the defection-prone $R<1-$region. In effect, individuals' (mis)learn what is best for them and end up contributing even though this would not be a unilateral best reply. Similarly, in the $R>1$ region free-riding starts to spread despite of the fact that it is obviously better to cooperate. For both these rather surprising and counterintuitive outcomes to emerge, the only thing needed is directional learning.}
\end{figure}

If $R<1$, the probability for any player to move from $p_i^t=1$ to $p_i^{t+T}=0 $ in some $T=1/\delta$ steps requires a single perturbation for that player and is therefore of the order of a single $\epsilon$. By contrast, the probability for any player to move from $p_i^t=0 $ to $ p_i^{t+T}=1$ in some $T$ steps is at least of the order $\epsilon^k$, because at least $k$ players (corresponding to the maximal $k$-strength of the equilibrium) must contribute in order for all of these players to experience a payoff increase. Notice that $k$ decreases in $R$. Again, the length of the path is $T=1/\delta$ steps, and that path requires no perturbations along the way, which is less likely the smaller $\delta$. With this, we conclude the proof of the proposition. However, it is also worth noting a direct corollary of the proposition; namely, as $\epsilon\rightarrow 0$, $E[p^t]\rightarrow 1$ if $ R\geq 1$, and $E[p^t]\rightarrow 0$ if $R<1$.

Lastly, we simulate the perturbed public goods game with directional learning and determine the actual average contribution levels in the stationary state. Color encoded results in dependence on the normalized rate of return $R$ and the responsiveness of players to the success of their past actions $\delta$ (alternatively, the sensitivity of the individual learning process) are presented in Fig.~\ref{simul} for $\epsilon=0.1$. Small values of $\delta$ lead to a close convergence to the respective Nash equilibrium of the game, regardless of the value of $R$. As the value of $\delta$ increases, the pure Nash equilibria erode and give way to a mixed outcome. It is important to emphasize that this is in agreement, or rather, this is in fact a consequence of the low $k-$strengths of the non-contribution pure equilibria (see Fig~\ref{ana}). Within intermediate to large $\delta$ values the Nash equilibria are implemented in a zonal rather than pinpoint way. When the Nash equilibrium is such that all players contribute ($R>1$), then small values of $\delta$ lead to more efficient aggregate play (recall any such equilibrium is $n-$strong). Conversely, by the same logic, when the Nash equilibrium is characterized by universal free-riding, then larger values of $\delta$ lead to more efficient aggregate play. Moreover, the precision of implementation also depends on the rate of return in the sense that uncoordinated deviations of groups of players lead to more efficient outcomes the higher the rate of return. In other words, the free-riding problem is mitigated if group deviations lead to higher payoffs for every member of an uncoordinated deviation group, the minimum size of which (that in turn is related to the maximal $k-$ strength of equilibrium) is decreasing with the rate of return.

\begin{figure}
\centerline{\epsfig{file=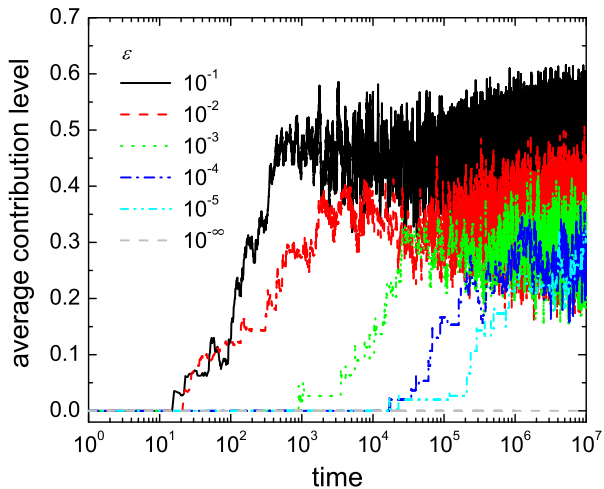,width=8cm}}
\caption{\label{eps} Time evolution of average contribution levels, as obtained for $R=0.7$, $\delta=0.1$ and different values of $\epsilon$ (see legend). If only $\epsilon>0$, the Nash equilibrium erodes to a stationary state where at least some members of the population always contribute to the common pool. There is a discontinuous transition to complete free-riding (defection) as $\epsilon \to 0$. Understandably, the lower the value of $\epsilon$ (the smaller the probability for the perturbation), the longer it may take for the drift to gain on momentum and for the initial deviation to evolve towards the mixed stationary state. Note that the time horizontally is in logarithmic scale.}
\end{figure}

Simulations also confirm that the evolutionary outcome is qualitatively invariant to: i) The value of $\epsilon$ as long as the latter is bounded away from zero, although longer convergence times are an inevitable consequence of very small $\epsilon$ values (see Fig.~\ref{eps}); ii) The replication of the population (i.e., making the whole population a group) and the random remixing between groups; and iii) The population size, although here again the convergence times are the shorter the smaller the population size. While both ii and iii are a direct consequence of the fact that we have considered the public goods game in a well-mixed rather than a structured population (where players would have a limited interaction range and where thus pattern formation could play a decisive role \cite{perc_jrsi13}), the qualitative invariance to the value of $\epsilon$ is elucidated further in Fig.~\ref{eps}. We would like to note that by ``qualitative invariance'' it is meant that, regardless of the value of $\epsilon>0$, the population always diverges away from the Nash equilibrium towards a stable mixed stationary state. But as can be observed in Fig.~\ref{eps}, the average contribution level and its variance both increase slightly as $\epsilon$ increases. This is reasonable if one considers $\epsilon$ as an exploration or mutation rate. More precisely, it can be observed that, the lower the value of $\epsilon$, the longer it takes for the population to move away from the Nash equilibrium where everybody contributes zero in the case that $1/n<R<1$ (which was also the initial condition for clarity). However, as soon as initial deviations (from $p_i=0$ in this case) emerge (with probability proportional to $\epsilon)$, the neutral rule in the original learning dynamics takes over, and this drives the population towards a stable mixed stationary state. Importantly, even if the value of $\epsilon$ is extremely small, the random drift sooner or later gains momentum and eventually yields similar contribution levels as those attainable with larger values of $\epsilon$. Most importantly, note that there is a discontinuous jump towards staying in the Nash equilibrium, which occurs only if $\epsilon$ is exactly zero. If $\epsilon$ is bounded away from zero, then the free-riding Nash equilibrium erodes unless it is $n-$strong (for very low values of $R\leq 1/n$).

\section*{Discussion}
We have introduced a public goods game with directional learning, and we have studied how the level of contributions to the common pool depends on the rate of return and the responsiveness of individuals to the successes and failures of their own past actions. We have shown that directional learning alone suffices to explain deviations from the Nash equilibrium in the stationary state of the public goods game. Even though players have no strategically relevant information about the game and/ or about each others' actions, the population could still end up in a mixed stationary state where some players contributed at least part of the time although the Nash equilibrium would be full free-riding. Vice versa, defectors emerged where cooperation was clearly the best strategy to play. We have explained these evolutionary outcomes by introducing the concept of $k-$strong equilibria, which bridge the gap between Nash and Aumann-strong equilibria. We have demonstrated that the lower the maximal $k-$strength and the higher the responsiveness of individuals to the consequences of their own past strategy choices, the more likely it is for the population to (mis)learn what is the objectively optimal unilateral (Nash) play.

These results have some rather exciting implications. Foremost, the fact that the provisioning of public goods even under adverse conditions can be explained without any sophisticated and often lengthy arguments involving selflessness or social preference holds promise of significant simplifications of the rationale behind seemingly irrational individual behavior in sizable groups. It is simply enough for a critical number (depending on the size of the group and the rate of return) of individuals to make a ``wrong choice'' at the same time once, and if only the learning process is sufficiently fast or naive, the whole subpopulation is likely to adopt this wrong choice as their own at least part of the time. In many real-world situations, where the rationality of decision making is often compromised due to stress, propaganda or peer pressure, such ``wrong choices'' are likely to proliferate. As we have shown in the context of public goods games, sometimes this means more prosocial behavior, but it can also mean more free-riding, depending only on the rate of return.

The power of directional (mis)learning to stabilize unilaterally suboptimal game play of course takes nothing away from the more traditional and established explanations, but it does bring to the table an interesting option that might be appealing in many real-life situations, also those that extend beyond the provisioning of public goods. Fashion trends or viral tweets and videos might all share a component of directional learning before acquiring mainstream success and recognition. We hope that our study will be inspirational for further research in this direction. The consideration of directional learning in structured populations \cite{szabo_pr07, perc_bs10}, for example, appears to be a particularly exciting future venture.

\section*{Methods}
For the characterization of the stationary states, we introduce the concept of $k-$strong equilibria, which nests both the Nash equilibrium \cite{nash_pnas50} and the Aumann-strong equilibrium \cite{aumann_74, aumann_87} as two special cases. While the Nash equilibrium describes the robustness of an outcome against unilateral ($1$-person) deviations, the Aumann-strong equilibrium describes the robustness of an outcome against the deviations of any subgroup of the population. An equilibrium is said to be (Aumann-)strong if it is robust against deviations of the whole population or indeed of any conceivable subgroup of the population, which is indeed rare. Our definition of the $k-$strong equilibrium bridges the two extreme cases, measuring the size of the group $k\geq 1$ (at or above Nash) and hence the degree to which an equilibrium is stable. We note that our concept is related to coalition-proof equilibrium \cite{bernheim_87, moreno_96}. In the public goods game, the free-riding Nash equilibrium is typically also more than $1-$strong but never $n-$strong. As we will show, the maximal strength $k$ of an equilibrium translates directly to the level of contributions in the stationary distribution of our process, which is additionally determined by the normalized rate of return $R$ and the responsiveness of players to the success of their past actions $\delta$, i.e., the sensitivity of the individual learning process.

\begin{acknowledgments}
This research was supported by the European Commission through the ERC Advanced Investigator Grant `Momentum' (Grant 324247), by the Slovenian Research Agency (Grant P5-0027), and by the Deanship of Scientific Research, King Abdulaziz University (Grant 76-130-35-HiCi).
\end{acknowledgments}

\end{document}